\documentclass[conf, letter, 10pt]{IEEEtran}
\usepackage[utf8]{inputenc}
\usepackage{amsmath}
\usepackage{amsfonts}
\usepackage{amssymb}
\usepackage{amsthm}
\usepackage{commath}
\usepackage{graphicx}
\usepackage{bm}
\usepackage{bbm}
\usepackage{multirow}
\usepackage{epstopdf}
\usepackage{physics}
\usepackage{mathtools}
\usepackage{cases}
\usepackage{enumitem}
\usepackage{cite}
\usepackage{xcolor}
\usepackage{stfloats}
\usepackage{float}
\usepackage{balance}
\usepackage{caption}
\usepackage{subcaption}
\allowdisplaybreaks
\newtheorem{theorem}{Theorem}
\newtheorem{definition}{Definition}
\newtheorem{lemma}{Lemma}
\newtheorem{proposition}{Proposition}
\newtheorem{corollary}{Corollary}

\theoremstyle{definition}

\theoremstyle{definition}
\newtheorem{example}{Example}
\theoremstyle{definition}

\newcommand\scalemath[2]{\scalebox{#1}{\mbox{\ensuremath{\displaystyle #2}}}}

\makeatletter
\renewcommand*\env@matrix[1][\arraystretch]{%
  \edef\arraystretch{#1}%
  \hskip -\arraycolsep
  \let\@ifnextchar\new@ifnextchar
  \array{*\c@MaxMatrixCols c}}
\makeatother

\newcommand{\mli}[1]{\mathit{#1}}
\begin{document}

\title{Adaptive Coding for Two-Way Lossy Source-Channel Communication\thanks{This work was supported in part by NSERC of Canada.}}

\author{\IEEEauthorblockN{Jian-Jia Weng, Fady Alajaji, and Tam\'as Linder}\\
\IEEEauthorblockA{Department of Mathematics and Statistics \\
Queen's University\\
Kingston, ON K7L 3N6, Canada \\
jian-jia.weng@queensu.ca, \{fady, linder\}@mast.queensu.ca}
}

\maketitle
\begin{abstract}
  An adaptive joint source-channel coding (JSCC) scheme is presented for transmitting correlated sources over discrete-memoryless two-way channels subject to distortion constraints. 
  The proposed JSCC scheme makes use of the previously transmitted and received channel signals as well as the sources' correlation to facilitate coordination between terminals. 
  It is shown that the adaptive scheme strictly subsumes prior lossy coding methods for two-way simultaneous transmission and yields a new adaptive separate source-channel coding result. 
  Two examples are given to show the scheme's advantages.  
\end{abstract}


\section{Introduction}\label{sec:introduction}
Shannon's two-way channel (TWC) \cite{shannon1961} enables full-duplex data transfer between two terminals. 
Ideally, each terminal causally generates its channel inputs by adapting them to previously received signals. 
However, as the two terminals are generally uncoordinated, the problem of how adaptive coding can provide reliable communication is not fully understood. 
The best-known results to date include Han's coding method \cite{han1984} for discrete-memoryless TWCs (DM-TWCs) and the Q-graph coding method for single-output DM-TWCs \cite{sabag2018}.

Beyond the channel coding problem, joint source-channel coding (JSCC) has recently received increased attention. 
The authors in \cite{maor2006} investigated an interactive lossy transmission scheme for sending correlated sources over two independent one-way channels. 
The two-way simultaneous transmission counterpart of that problem was considered in \cite{gunduz2009,jjw2017,jjw2019isit} for general DM-TWCs.
The (non-adaptive) two-way hybrid coding scheme of \cite{jjw2019isit} subsumes all prior results in this setup, but it does not use adaptive coding. 

Using the adaptive channel coding idea of \cite{han1984},\footnote{We remark that the two-way source coding results of \cite{kaspi1985} and \cite{permuter2010} can also be used to design adaptive JSCC schemes, but the details are not covered here due to length constraints.} we devise an adaptive JSCC scheme for the two-way lossy simultaneous transmission.
Roughly speaking, we couple the two terminals' encoding and transmission processes through a stationary Markov chain. 
Although the terminals operate independently, the adaptive encoding procedure driven by the Markov chain ultimately coordinates their encoding operations, thus jointly optimizing their transmissions.
Our proposed adaptive JSCC scheme not only strictly generalizes the two-way hybrid coding scheme of \cite{jjw2019isit} (and hence all its special cases).
It also yields a new adaptive separate source-channel coding (SSCC) scheme which consists of the concatenation of Wyner-Ziv (WZ) source coding \cite{wyner1976} and Han's adaptive channel coding \cite{han1984}. 

The rest of this paper is organized as follows. 
In Section~II, the system model and definitions are introduced.  
Our achievability result is presented in Section~III; its proof is relegated to the Appendix. 
Special cases and examples are given in Section~IV, and conclusions are drawn in Section~V. 

\section{Preliminaries}\label{sec:p2p}
For any $l\ge 1$, let $A^l\triangleq (A_1, A_2, \dots, A_l)$ denote a length-$l$ sequence of random variables with common alphabet $\mathcal{A}$.
The realization of $A^l$ will be denoted by $a^{l}=(a_1, a_2, \dots, a_l)\in\mathcal{A}^l$. 
In the paper, all alphabets are assumed to be finite. 

\begin{figure}[!t]
  \centering
  \includegraphics[draft=false, scale=0.435]{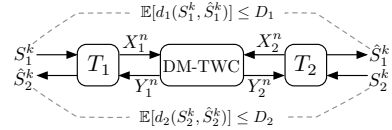}
  \caption{The block diagram for the lossy transmission of correlated source $(S_1^k, S_2^k)$ via $n$ uses of a DM-TWC.}
  \label{fig:TWCblcok}
  \vspace{-0.4cm}
\end{figure}

As depicted in Fig.~\ref{fig:TWCblcok}, two terminals exchange correlated source messages $S_1^k$ and $S_2^k$ via $n$ channel uses subject to distortion constraints, where $n, k\in\mathbb{Z}_{+}$.
The source pair $(S_1^k, S_2^k)$ is stationary and memoryless in time having the common joint probability distribution $P_{S_1, S_2}$, i.e., $P_{S_1^k, S_2^k}(s_1^k, s_2^k)=\prod_{m=1}^k P_{S_1, S_2}(s_{1, m}, s_{2, m})$, where $(s_{1, m}, s_{2, m})\in\mathcal{S}_1\times\mathcal{S}_2$. 
For $j=1, 2$, the reconstruction $\hat{s}_j^k$ of a given source message $s_j^k$ is assessed by $d_j(s_j^k, \hat{s}_j^k)\,\triangleq\,k^{-1}\sum_{m=1}^k d_j({s_{j, m}, \hat{s}_{j, m}})$, where $d_j: \mathcal{S}_j\times\mathcal{\hat{S}}_j{\rightarrow}\mathbb{R}_{+}$ is a single-letter distortion measure.
Let $X_{j, i}$ and $Y_{j, i}$ denote the channel input and output of terminal~$j$ at time $i$, respectively. 
We consider a DM-TWC with transition probability $P_{Y_1, Y_2|X_1, X_2}$. 
A joint source-channel code in this problem setup is defined as follows.  

\begin{definition}
An $(n, k)$ code for transmitting $(S_1^k, S_2^k)$ over a DM-TWC consists of two sequences of encoding functions $f_1\triangleq \{f_{1, i}\}_{i=1}^n$ and $f_2\triangleq \{f_{2, i}\}_{i=1}^n$ such that $X_{1,1}= f_{1, 1}(S_1^k)$, $X_{2,1}=f_{2, 1}(S_2^k)$, $X_{1,i}=f_{1, i}(S_1^k, Y_1^{i-1})$, and $X_{2,i}=f_{2, i}(S_2^k, Y_2^{i-1})$ for $i=2, 3, \dots, n$, and two decoding functions $g_1$ and $g_2$ such that $\hat{S}_2^k=g_1(S_1^k, Y_1^n)$ and $\hat{S}_1^k=g_2(S_2^k, Y_2^n)$.
\end{definition}

The rate of the above joint source-channel code is given by $k/n$ (source symbols/channel use), and the associated expected distortion is $D_j(k)=\mathbb{E}[d_j(S_j^k, \hat{S}_j^k)]$ for $j=1, 2$, where the expectation is taken with respect to the joint distribution
\begin{IEEEeqnarray}{l}
  \scalemath{1}{P_{S_1^k, S_2^k, X_1^n, X_2^n, Y_1^n, Y_2^n}=P_{S_1^k, S_2^k}\Bigg(\prod\limits_{i=1}^n P_{X_{1, i}|S_1^k, Y_1^{i-1}}\Bigg)}\nonumber\allowdisplaybreaks\\
\ \ \qquad\qquad\scalemath{1}{\Bigg(\prod\limits_{i=1}^n P_{X_{2, i}|S_2^k, Y_2^{i-1}}\Bigg)\Bigg(\prod\limits_{i=1}^n P_{Y_{1,i}, Y_{2, i}|X_{1, i}, X_{2, i}}\Bigg)},\nonumber
\end{IEEEeqnarray}
where $P_{Y_{1,i}Y_{2,i}|X_{1,i},X_{2,i}}=P_{Y_1,Y_2|X_1,X_2}$ for $i=1,\ldots,n$.

\begin{definition}
A distortion pair $(D_1, D_2)$ is said to be achievable at rate $R$ over a DM-TWC if there exists a sequence of $(n, k)$ joint source-channel codes (where $n$ is a function of $k$) such that for $\lim_{k\to\infty} k/n = R$ and $\limsup_{k\to\infty}\allowbreak D_j(k)\le D_j$ $j=1, 2$. 
The achievable distortion region of a rate $R$ two-way lossy transmission system is defined as the convex closure of the set of all achievable distortion pairs at rate $R$. 
\end{definition}

Prior achievability results for this problem mainly involve non-adaptive JSCC coding, i.e., coding schemes where $X_{j, i}=f_{j, i}(S_j^k)$ for all~$i$ \cite{gunduz2009,jjw2017,jjw2019isit}. 
In some special cases, such schemes are optimal.
We now review the most general of these results, which is derived from the hybrid coding scheme of \cite{kim2015}. 
As $P_{S_1, S_2}$ and $P_{Y_1, Y_2|X_1, X_2}$ are fixed and given by the problem setup, we will not refer to them in the result statements. 

\begin{proposition}[\emph{Two-Way Hybrid Coding Scheme}, \cite{jjw2019isit}]
  \label{prop:hybrid}
  A distortion pair $(D_1, D_2)$ is achievable for the rate-one ($R=1$) lossy transmission of correlated sources over a DM-TWC if
  \begin{subequations}
  \label{eq:hybrid}
  \begin{IEEEeqnarray}{rCl}
  I(S_1; U_1|S_2, U_2) < I(U_1; Y_2|S_2, U_2),\label{eq:hybrida}\\
  I(S_2; U_2|S_1, U_1) < I(U_2; Y_1|S_1, U_1),\label{eq:hybridb}
  \end{IEEEeqnarray}
  \end{subequations}
  where $P_{U_1, U_2|S_1, S_2}=P_{U_1|S_1}P_{U_2|S_2}$ for some $P_{U_1|S_1}$ and $P_{U_2|S_2}$, and there exist encoding functions $X_j=f_j(S_j, U_j)$ and decoding functions $\hat{S}_{j'}=g_j(U_{j'}, S_j, U_j, Y_j)$ such that $\mathbb{E}[d_j(S_j, \hat{S}_j)]\le D_j$ for $j, j'=1, 2$ with $j\neq j'$.
\end{proposition}

\begin{figure}[t!]
  \centering
  \includegraphics[draft=false, scale=0.475]{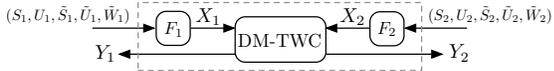}
  \caption{An illustration of two-way coded channel.}
  \label{fig:codedTWC}
  \vspace{-0.3cm}
\end{figure}

\begin{figure*}[b!]
  \vspace{-0.4cm}
  \normalsize
  \hrulefill
  \begin{IEEEeqnarray}{l}
    P_{Y_1, Y_2|S_1, S_2, U_1, U_2, \tilde{S}_1, \tilde{S}_2, \tilde{U}_1, \tilde{U}_1, \tilde{W}_1, \tilde{W}_2}(y_1, y_2|s_1, s_2, u_1, u_2, \tilde{s}_1, \tilde{s}_2, \tilde{u}_1, \tilde{u}_1, \tilde{w}_1, \tilde{w}_2)\nonumber\\
    \quad =\sum_{x_1, x_2} \mathbbm{1}\{x_1=F_1(s_1, u_1, \tilde{s}_1, \tilde{u}_1, \tilde{w}_1)\}\mathbbm{1}\{x_2=F_2(s_2, u_2, \tilde{s}_2, \tilde{u}_2, \tilde{w}_2)\}P_{Y_1, Y_2|X_1, X_2}(y_1, y_2|x_1, x_2).\label{eq:sysprob}\IEEEeqnarraynumspace\\
    P_{Z^{(t)}|Z^{(t-1)}}(s_1, s_2, u_1, u_2, \tilde{s}_1, \tilde{s}_2, \tilde{u}_1, \tilde{u}_2, \tilde{w}_1, \tilde{w}_2, x_1, x_2, y_1, y_2|s'_1, s'_2, u'_1, u'_2, \tilde{s}'_1, \tilde{s}'_2, \tilde{u}'_1, \tilde{u}'_2, \tilde{w}'_1, \tilde{w}'_2, x'_1, x'_2, y'_1, y'_2)\nonumber\\
    \quad = P_{S_1, S_2}(s_1, s_2)P_{U_1|S_1}(u_1|s_1)P_{U_2|S_2}(u_2|s_2)\mathbbm{1}\{\tilde{s}_1=s'_1\}\mathbbm{1}\{\tilde{s}_2=s'_2\}\mathbbm{1}\{\tilde{u}_1=u'_1\}\mathbbm{1}\{\tilde{u}_2=u'_2\}\mathbbm{1}\{\tilde{w}_1=(x'_1, y'_1)\}\nonumber\\
    \qquad\quad\mathbbm{1}\{\tilde{w}_2=(x'_2, y'_2)\}\mathbbm{1}\{x_1=F_1(s_1, u_1, \tilde{s}_1, \tilde{u}_1, \tilde{w}_1)\}\mathbbm{1}\{x_2=F_2(s_2, u_2, \tilde{s}_2, \tilde{u}_2, \tilde{w}_2)\}P_{Y_1, Y_2|X_1, X_2}(y_1, y_2|x_1, x_2).\label{eq:trankernel}\IEEEeqnarraynumspace
  \end{IEEEeqnarray}
\end{figure*}

\vspace{-0.3cm}
\section{An Adaptive Two-Way Lossy JSCC Scheme}
Inspired by Han's work \cite{han1984}, we propose to integrate adaptive hybrid coding in a Markovian transmission framework to exploit the advantages of both methods. 
Specifically, we modify Han's channel coding method \cite{han1984} for the use of JSCC.
Without loss of generality, we only consider rate-one transmission $(n=k)$.
In brief, instead of exchanging a single block of memoryless source messages $(S_1^n, S_2^n)$ via $n$ channel uses, we exchange $B$ blocks of such source messages via $n(B+1)$ channel uses for some $B\ge 1$. 
The extra $n$ channel uses can be viewed as added redundancy for data protection. 
We next adopt the superposition Markov coding framework of \cite{han1984}: each block of source messages is encoded across two consecutive transmission blocks and is decoded at the end of the second block.
However, within each transmission block, an adaptive hybrid JSCC coding scheme, an extension of its non-adaptive counterpart in \cite{jjw2019isit}, is employed. 
We point out that the above modifications and the ensuing derivation of our main achievability result (in Theorem~1) are intricate.
We next describe the key technical ingredients used in obtaining the result (whose proof is sketched in the Appendix).
%
\vspace{-0.3cm}
\subsection{Two-Way Coded Channels}
Consider an auxiliary coded channel built on the original physical DM-TWC, as shown in Fig.~\ref{fig:codedTWC}. 
The coded channel has inputs $S_j, U_j, \tilde{S}_j, \tilde{U}_j$ and $\tilde{W}_j$ at terminal~$j=1, 2$. 
The inputs $(S_j, U_j)$ and $(\tilde{S}_j, \tilde{U}_j)$ will be used to carry new and old source information, respectively, where $U_j$ (resp. $\tilde{U}_j$) denotes the coded version of $S_j$ (resp. $\tilde{S}_j$); the input $\tilde{W}_j$ represents the past channel inputs and outputs at terminal~$j$. 
The new channel also involves two encoding functions $F_j: \mathcal{S}_j\times \mathcal{U}_j\times\tilde{\mathcal{S}}_j\times\tilde{\mathcal{U}}_j\times\tilde{\mathcal{W}}_j\to\mathcal{X}_j$, which transform the inputs of the coded channel into the inputs of the original DM-TWC.
The outputs of the new channel are still $Y_1$ and $Y_2$. 
The joint probability distribution of the channel inputs is given as
$P_{S_1, S_2, U_1, U_2, \tilde{S}_1, \tilde{S}_2, \tilde{U}_1, \tilde{U}_2, \tilde{W}_1, \tilde{W}_2}= P_{S_1, S_2}P_{U_1|S_1}P_{U_2|S_2}P_{\tilde{S}_1, \tilde{S}_2, \tilde{U}_1, \tilde{U}_2, \tilde{W}_1, \tilde{W}_1}.$
Here, the parameters $P_{U_1|S_1}$, $P_{U_2|S_2}$, $P_{\tilde{S}_1, \tilde{S}_2, \tilde{U}_1, \tilde{U}_2, \tilde{W}_1, \tilde{W}_1}$, and $F_j$ are part of our JSCC code.
The transition probability of the coded channel is given in \eqref{eq:sysprob}, where $\mathbbm{1}\{\cdot\}$ denotes the indicator function.
We remark that $\tilde{W}_j$ in $F_j$ enables adaptive coding for the channel inputs while $(\tilde{S}_j, \tilde{U}_j)$ are retained for superposition coding.

\vspace{-0.3cm}
\subsection{Markov Chain for the Coded Channel}
Given a configuration $\{P_{U_1|S_1}, P_{U_2|S_2}, \allowbreak P_{\tilde{S}_1, \tilde{S}_2, \tilde{U}_1, \tilde{U}_2, \tilde{W}_1, \tilde{W}_2},\allowbreak F_1, \allowbreak F_2\}$ for the two-way coded channel, we next construct a time-homogeneous discrete-time Markov chain for the overall system with state space: $\mathcal{S}_1\times\mathcal{S}_2\times \mathcal{U}_1\times \mathcal{U}_2\times \tilde{\mathcal{S}}_1\times \tilde{\mathcal{S}}_2\times \tilde{\mathcal{U}}_1\times \mathcal{\tilde{U}}_2\times\tilde{\mathcal{W}}_1\times \tilde{\mathcal{W}}_2\times\mathcal{X}_1\times\mathcal{X}_2\times\mathcal{Y}_1\times\mathcal{Y}_2$, where $\tilde{\mathcal{S}}_j\triangleq\mathcal{S}_j$, $\tilde{\mathcal{U}}_j\triangleq\mathcal{U}_j$, and $\tilde{\mathcal{W}}_j\triangleq\mathcal{X}_j\times\mathcal{Y}_j$ for $j=1, 2$. 
Let $Z^{(t)}\triangleq(S^{(t)}_1, \allowbreak S^{(t)}_2, \allowbreak U^{(t)}_1, \allowbreak U^{(t)}_2, \allowbreak\tilde{S}^{(t)}_1, \allowbreak\tilde{S}^{(t)}_2, \allowbreak\tilde{U}^{(t)}_1, \allowbreak\tilde{U}^{(t)}_2, \allowbreak\tilde{W}^{(t)}_1, \allowbreak\tilde{W}^{(t)}_2, X^{(t)}_1, \allowbreak X^{(t)}_2, \allowbreak Y^{(t)}_1, \allowbreak Y^{(t)}_2)$ denote the state of the Markov chain at time $t$, where $\tilde{S}^{(t)}_j\triangleq S^{(t-1)}_j$, $\tilde{U}^{(t)}_j\triangleq U^{(t-1)}_j$, $\tilde{W}^{(t)}_j\triangleq \allowbreak(X^{(t-1)}_j, Y^{(t-1)}_j)$, and $(\tilde{S}^{(1)}_1, \allowbreak\tilde{S}^{(1)}_2, \allowbreak\tilde{U}^{(1)}_1, \allowbreak\tilde{U}^{(1)}_2, \allowbreak\tilde{W}^{(1)}_1, \allowbreak\tilde{W}^{(1)}_2)$ is initialized according to $P_{\tilde{S}_1, \tilde{S}_2, \tilde{U}_1, \tilde{U}_2, \tilde{W}_1, \tilde{W}_2}$. 
Moreover, the quadruple $(S^{(t)}_1, S^{(t)}_2, U^{(t)}_1, U^{(t)}_2)$ is generated according to $P_{S_1, S_2, U_1, U_2}= \allowbreak P_{S_1, S_2}\allowbreak  P_{U_1|S_1}\allowbreak P_{U_2|S_2}$ independent of $(\tilde{S}^{(t)}_1, \allowbreak\tilde{S}^{(t)}_2, \allowbreak\tilde{U}^{(t)}_1, \allowbreak\tilde{U}^{(t)}_2, \allowbreak\tilde{W}^{(t)}_1, \allowbreak\tilde{W}^{(t)}_2)$. 
The physical channel inputs are naturally produced as $X^{(t)}_j=F_j(S_j^{(t)}, U_j^{(t)}, \tilde{S}_j^{(t)}, \tilde{U}_j^{(t)}, \tilde{W}_j^{(t)})$ and the corresponding channel outputs are $Y^{(t)}_j$, $j=1, 2$. 
Based on this construction, the transition kernel is obtained in \eqref{eq:trankernel} for $t\ge 2$. 
One can readily verify that the process $\{Z^{(t)}\}$ is a first-order time-homogeneous Markov chain. 
Note that the chain may not be stationary for a specific configuration. 

\vspace{-0.2cm}
\subsection{Stationary Distribution under Distortion Constraints} 
To obtain an achievability result with time-independent conditions, we only consider  configurations that induce a stationary Markov chain.  
We remark that given any fixed $P_{U_j|S_j}$'s and $F_j$'s, one can always find a $P_{\tilde{S}_1, \tilde{S}_2, \tilde{U}_1, \tilde{U}_2, \tilde{W}_1, \tilde{W}_2}$ that induces a stationary Markov chain based on the simplified condition $P_{\tilde{S}_1^{(1)}, \tilde{S}_2^{(1)}, \tilde{U}_1^{(1)}, \tilde{U}_2^{(1)}, \tilde{W}_1^{(1)}, \tilde{W}_2^{(1)}}(\tilde{s}_1, \tilde{s}_2, \tilde{u}_1, \tilde{u}_2, \tilde{w}_1, \tilde{w}_2)=\allowbreak P_{\tilde{S}_1^{(2)},\tilde{S}_2^{(2)}, \tilde{U}_1^{(2)}, \tilde{U}_2^{(2)}, \tilde{W}_1^{(2)}, \tilde{W}_2^{(2)}}(\tilde{s}_1, \tilde{s}_2, \tilde{u}_1, \tilde{u}_2, \tilde{w}_1, \tilde{w}_2)$ for all $\tilde{s}_1$, $\tilde{s}_2$, $\tilde{u}_1$, $\tilde{u}_2$, $\tilde{w}_1$, and $\tilde{w}_2$. 
Such a $P_{\tilde{S}_1, \tilde{S}_2, \tilde{U}_1, \tilde{U}_2, \tilde{W}_1, \tilde{W}_1}$ is guaranteed to exist since the Markov chain has finite state space. 
For the source reconstruction, we next associate the configuration with the functions $G_j: \tilde{\mathcal{U}}_{j'}\times\mathcal{S}_j\times \mathcal{U}_j \times\tilde{\mathcal{S}}_j\times\tilde{\mathcal{U}}_j\times\tilde{\mathcal{W}}_j\times {\mathcal{Y}}_j\to\hat{\mathcal{S}}_{j'}$ for $j, j'=1, 2$ with $j\neq j'$.
As will be seen later (in the Appendix) in our coding scheme, terminal~$j$ first decodes $\tilde{U}_{j'}$ and then reconstruct $S_{j'}$ via $G_j$. 
Let $\Pi_{Z}(D_1, D_2)$ denote the set of all configurations that induce a stationary chain and satisfy the distortion constraints: $\mathbb{E}[d_j(S_j, \hat{S}_j)]\le D_j$ for $j=1, 2$.
Note that $\Pi_{Z}$ might be empty for some $(D_1, D_2)$. \smallskip

Based on the above setup,\footnote{
Our approach relies on the superposition coding idea of \cite{cover1975bc}, which facilitates the derivations of special cases from our main theorem. 
It is possible to reduce the number of auxiliary random variables and/or simplify their joint probability distribution using other coding methods such as \cite{bergmans1973dbc}.  
} we are ready to present our main result. Note that the associated coding scheme and proof sketch are given in the Appendix.
In Theorem~\ref{thm:main} (and the special cases in Section~IV), one can convexify the achievable distortion region via a standard time-sharing argument.

\begin{theorem}
  \label{thm:main}
  A distortion pair $(D_1, D_2)$ is achievable for the rate-one lossy transmission of correlated sources over a DM-TWC if there exists a configuration in $\Pi_{Z}(D_1, D_2)$ such that
  \begin{IEEEeqnarray}{rCl}
    \label{eq:mainconds}
    \IEEEyesnumber
    \IEEEyessubnumber*
    I(\tilde{S}_1; \tilde{U}_1)&< & I(\tilde{U}_1; S_2, U_2, \tilde{S}_2, \tilde{U}_2, \tilde{W}_2, X_2, Y_2),\label{eq:mainconda}\\
    I(\tilde{S}_2; \tilde{U}_2)&< & I(\tilde{U}_2; S_1, U_1, \tilde{S}_1, \tilde{U}_1, \tilde{W}_1, X_1, Y_1).\label{eq:maincondb}
  \end{IEEEeqnarray}
\end{theorem}

\section{Further Exploration of the JSCC Scheme}
This section illustrates two special cases of Theorem~\ref{thm:main}. 
Two examples are also given to reveal the generality of the theorem.

\vspace{-0.35cm}
\subsection{Special Cases}
\subsubsection{Two-Way Hybrid Coding Scheme \cite{jjw2019isit}} 
Choose $(P_{U_j|S_j},\allowbreak f_j, \allowbreak g_j)$, $j=1, 2$, in Proposition~\ref{prop:hybrid} that attain the distortion pair $(D_1, D_2)$. 
In Theorem~\ref{thm:main}, we let $F_j(s_j, \allowbreak u_j, \allowbreak \tilde{s}_j, \allowbreak \tilde{u}_j, \allowbreak \tilde{w}_j)=f_j(\tilde{s}_j, \allowbreak \tilde{u}_j)$ and $G_j(\tilde{u}_{j'}, s_j, u_j, \tilde{s}_j, \tilde{u}_j, \tilde{w}_j, y_j)=g_j(\tilde{u}_{j'}, \allowbreak \tilde{s}_j, \allowbreak \tilde{u}_j, \allowbreak y_j)$. 
As noted in the previous section, for any given $(P_{U_j|S_j}, f_j, g_j)$, $j=1, 2$, there exists at least one stationary distribution for $\{Z^{(t)}\}$. 
Furthermore, our construction of $\{Z^{(t)}\}$ ensures that the marginal distribution $P_{\tilde{S}_1, \tilde{S}_2, \tilde{U}_1, \tilde{U}_2, X_1, X_2, Y_1, Y_2}$ of each stationary distribution is identical to the joint distribution $P_{S_1, S_2, U_1, U_2, X_1, X_2, Y_1, Y_2}$ specified in Proposition~\ref{prop:hybrid}, thus satisfying the same distortion constraint, i.e., the above configuration is in $\Pi_{Z}(D_1, D_2)$. 
Next, observing that $\tilde{W}_j$ is independent of $(\tilde{S}_{j'}, S_j, U_j, \tilde{S}_j, \tilde{U}_j, X_j, Y_j)$ for $j=1, 2$ and using the fact that $\tilde{U}_j$ is independent of $(S_j, U_j)$, we can remove $(S_j, U_j, \tilde{W}_j)$ from \eqref{eq:mainconds} without changing the values on the right-hand-side of \eqref{eq:mainconds}. 
A simplification of \eqref{eq:mainconds} further results in
\vspace{-0.1cm}
\begin{IEEEeqnarray}{rCl}
\label{eq:special1}
\IEEEyesnumber
\IEEEyessubnumber*
I(\tilde{S}_1; \tilde{U}_1|\tilde{S}_2, \tilde{U}_2)&< & I(\tilde{U}_1; Y_2|\tilde{S}_2, \tilde{U}_2),\\
I(\tilde{S}_2; \tilde{U}_2|\tilde{S}_1, \tilde{U}_1)&< & I(\tilde{U}_2; Y_1|\tilde{S}_1, \tilde{U}_1).
\vspace{-0.1cm}  
\end{IEEEeqnarray}
Since the random variables in \eqref{eq:special1} have the same common distribution as their counterparts in \eqref{eq:hybrid}, the inequalities in \eqref{eq:special1} are identical to those in \eqref{eq:hybrid} and so Proposition~\ref{prop:hybrid} is recovered.\vspace{+0.2cm}


\subsubsection {The Concatenation of WZ Source Coding \cite{wyner1976} and Han's Adaptive Channel Coding \cite{han1984}} For $j\neq j'$, let $R^{(j)}_{\text{WZ}}(D_j)$ denote the WZ rate-distortion function of $S_j$ with auxiliary random variable $T_j$ and decoding function $h_{j'}: \mathcal{S}_{j'}\times\mathcal{T}_j\to\mathcal{S}_j$ such that $P_{S_j, S_{j'}, T_j}=P_{S_j, S_{j'}}P_{T_j|S_j}$ and $\mathbb{E}[d_j(S_j, h_{j'}(S_{j'}, T_j)]\le D_j$ \cite{wyner1976}. 
Furthermore, let $V_j$ and $\tilde{V}_j$ denote the auxiliary random variables used in Han's result \cite{han1984} and let $\gamma_j: \mathcal{V}_j\times\tilde{\mathcal{V}}_j\times \mathcal{W}_j\to\mathcal{X}_j$ denote terminal $j$'s encoding function. 
Also, assume that $\gamma_j$'s and $P_{V_1, V_2, \tilde{V}_1, \tilde{V}_2, \tilde{W}_1, \tilde{W}_2}\triangleq P_{V_1}P_{V_2}P_{\tilde{V}_1, \tilde{V}_2, \tilde{W}_1, \tilde{W}_2}$ induce a stationary Markov chain in Han's coding scheme. 
For $j, j'=1, 2$ with $j\neq j'$, we set $U_j=(T_j, V_j)$ and consider the following settings in Theorem~\ref{thm:main}: $F_j(s_j, \allowbreak u_j, \allowbreak \tilde{s}_j, \allowbreak \tilde{u}_j, \allowbreak \tilde{w}_j)=\gamma_j(v_j, \tilde{v}_j, \allowbreak \tilde{w}_j)$ and $G_j(\tilde{u}_{j'}, s_j, u_j, \tilde{s}_j, \tilde{u}_j, \tilde{w}_j, y_j)=h_{j}(\tilde{s}_j, \tilde{t}_{j'})$, $P_{U_j|S_j}=P_{T_j|S_j}P_{V_j}$, and $P_{\tilde{S}_1, \tilde{S}_2, \tilde{U}_1, \tilde{U}_2, \tilde{W}_1, \tilde{W}_1}=P_{\tilde{S}_1, \tilde{S}_2}P_{\tilde{T}_1|\tilde{S}_1}P_{\tilde{T}_2|\tilde{S}_2}P_{\tilde{V}_1, \tilde{V}_2, \tilde{W}_1, \tilde{W}_2}$. 
The configuration can be shown to induce a stationary Markov chain $\{Z^{(t)}\}$ and satisfy the distortion constraints $\mathbb{E}[d_j(S_j, h_{j'}(S_{j'}, T_j)]\le D_j$, $j=1, 2$. 
Also, the inequalities in \eqref{eq:mainconds} can be further simplified as in \eqref{eq:ssccwzhan} below, i.e., the WZ source coding rates are smaller than the channel coding rates of Han's coding scheme. 
We summarize this new SSCC result in the following corollary.

\begin{corollary}
  \label{cor:ssccwzhan}
  A distortion pair $(D_1, D_2)$ is achievable for the rate-one lossy transmission of correlated sources over a DM-TWC if
  \vspace{-0.35cm}
  \begin{subequations}
    \label{eq:ssccwzhan}
    \begin{IEEEeqnarray}{rCl}
      R^{(1)}_{\text{WZ}}(D_1)&< & I(\tilde{V}_1; X_2, Y_2, \tilde{V}_2, \tilde{W}_2),\label{eq:ssccwzhana}\\
      R^{(2)}_{\text{WZ}}(D_2)&< & I(\tilde{V}_2; X_1, Y_1, \tilde{V}_1, \tilde{W}_1),\label{eq:ssccwzhanb}
    \end{IEEEeqnarray}
    \end{subequations}
  for some joint probability distribution $P_{\tilde{V}_1, \tilde{V}_2, \tilde{W}_1, \tilde{W}_2, X_1, X_2}$ as defined in \cite[Section IV]{han1984}.
\end{corollary}

\vspace{-0.3cm}
\subsection{Examples}
Examples~1 and~2 below show that Theorem~\ref{thm:main} strictly generalizes Proposition~\ref{prop:hybrid} and Corollary~\ref{cor:ssccwzhan}, respectively. 
In both examples, the Hamming distortion is considered. 
We will need the following specialized converse result in Example~1.

\begin{lemma}[A special case of {\cite[Lemma 2]{jjw2017}}]
  \label{lma:converse1}
  Assume that the non-adaptive encoder $f_j: \mathcal{S}^{n}_j\to\mathcal{X}^{n}_j$ is used for $j=1, 2$.   
  If a distortion pair $(D_1, D_2)$ is achievable for the rate-one lossy transmission of independent sources over a DM-TWC, then\newpage
  \begin{IEEEeqnarray}{rCl}
  R^{(1)}(D_1)&\le & I(X_1; Y_2|X_2, Q),\nonumber\\
  R^{(2)}(D_2)&\le & I(X_2; Y_1|X_1, Q),\nonumber
  \end{IEEEeqnarray}
  for some $P_{Q, X_1, X_2}=P_{Q}P_{X_1|Q}P_{X_2|Q}$, where $R^{(j)}(D_j)$ denotes the standard rate-distortion function of $S_j$, $j=1, 2$.
\end{lemma}

\begin{example}[\bf{Sending Independent Binary Sources over Dueck's DM-TWC} \cite{dueck1979}]
  Let Ber$(p)$ denote a Bernoulli random variable with probability of success $p\in[0, 1]$. 
  Consider the independent sources $S_1=\text{Ber}(0.89)$ and $S_2=\text{Ber}(0.89)$ so that $H(S_1)=H(S_2)\approx 0.5$.
  We recall Dueck's DM-TWC \cite{dueck1979}, where $\bm{X}_j=(X_{j, 1}, X_{j, 2})$,\footnote{As Dueck's DM-TWC has $\mathcal{X}_j=\{0, 1\}^2$ and $\mathcal{Y}_j=\{0, 1\}^3$, we here use $(X_{j, 1}, X_{j, 2})\in\mathcal{X}_j$ to denote the two channel inputs of terminal $j$.} $\bm{Y}_j=(X_{1, 1}\cdot X_{2, 1}, N_j\oplus_2 X_{j', 2}, N_{j'})$ for $j, j'=1, 2$ with $j\neq j'$, $\oplus_2$ denotes the modulo-2 addition, and $N_1=\text{Ber}(0.5)$ and $N_2=\text{Ber}(0.5)$ are independent channel noises that are independent of all channel inputs and sources.
  Han showed in \cite{han1984} that the rate pair $(\tilde{R}_{\text{C}, 1}, \tilde{R}_{\text{C}, 2})=(0.5, 0.5)$ is not achievable via Shannon's random coding scheme but can be achieved via his adaptive channel coding scheme. 
  Based on this fact and Lemma~\ref{lma:converse1}, we conclude that the two-way hybrid coding scheme cannot achieve the  distortion pair $(D_1, D_2)=(0, 0)$ (since it uses non-adaptive encoders).
  By contrast, Corollary~\ref{cor:ssccwzhan} shows that it is achievable via our adaptive scheme as $R_{\text{WZ}, j}(0)=H(S_j)< \tilde{R}_{\text{C}, j}$ holds for $j=1, 2$. 
\end{example}

\begin{example}[\bf{Sending Correlated Binary Sources over Binary-Multiplying TWCs} \cite{shannon1961}]
  Consider the binary multiplying TWC given by $Y_j=X_1\cdot X_{2}$ for $j=1, 2$.
  The capacity region of the channel is not known, but it is known that any symmetric achievable channel coding rate pair is component-wise upper bounded by $(0.646, 0.646)$ \cite{hekstra1989}. 
  Suppose that we want to transmit binary correlated sources with joint probability distribution $P_{S_1, S_2}(0, 0)=0$ and $P_{S_1, S_2}(s_1, s_2)=1/3$ for $(s_1, s_2)\neq (0, 0)$. 
  The WZ coding theorem implies that the minimum source coding rate pair is $(H(S_1|S_2), H(S_2|S_1))=(0.667, 0.667)$. 
  Clearly, $(D_1, D_2)=(0, 0)$ is not achievable by \emph{any} SSCC scheme, including the adaptive coding scheme of Corollary~\ref{cor:ssccwzhan}, because the data compression rate exceeds the largest possible transmission rate for reliable communication. 
  However, the uncoded scheme: $X_j=S_j$ for $j=1, 2$ can be easily shown to provide lossless transmission. 
  As the adaptive coding scheme of Corollary~\ref{cor:ssccwzhan} and the uncoded scheme are special cases of our scheme, the result of Theorem~\ref{thm:main} strictly subsumes the result of Corollary~\ref{cor:ssccwzhan}. 
\end{example}


\section{Conclusions}
We generalized prior JSCC schemes for lossy two-way simultaneous transmission. 
Our adaptive coding result not only demonstrates a way to coordinate the two terminals' transmission but also underscores the importance of preserving source correlation. 
Although our scheme enlarges the achievable distortion region, its potential use in practice needs further study due to its high coding complexity. 
Directions to address this issue include the study of adaptive coding based on the SSCC structure and symbol-wise adaptive coding (as opposed to block-wise adaptive coding). 

\begin{appendix}[Proof of Theorem~\ref{thm:main}]
Let $\mathcal{T}_{\epsilon}^{(n)}$ denote the typical set with parameters $n$ and $\epsilon$ as defined in \cite{kim2011}; the domain of the sequences in $\mathcal{T}_{\epsilon}^{(n)}$ should be clear from the context and hence is omitted for the sake of brevity. 
For $j=1, 2$ and $b=1, 2, \cdots, B$, we define $2^{\mli{nR}^{(b)}_j}$ as the size of terminal~$j$'s codebook $\mathcal{C}^{(b)}_j$, which is used to encode the $b$-th block $\bm{S}^{(b)}_j$ of source messages.
Choose $\epsilon>\epsilon_1>0$. 
If $\mathcal{E}$ is an event, we let $\overline{\mathcal{E}}$ denote its complement. 
\smallskip

\begin{figure*}[b]
  \normalsize
  \hrulefill
  \begin{IEEEeqnarray}{rCl}
    \label{eq:errorevent}
    \IEEEyesnumber
    \IEEEyessubnumber*
    \scalemath{0.92}{\mathcal{E}_1^{(1)}}&\triangleq &\scalemath{0.92}{\{(\bm{S}_1^{(1)}, \bm{S}_2^{(1)}, \bm{U}_1^{(1)}(M_1^{(1)}), \bm{U}_2^{(1)}(M_2^{(1)}), \tilde{\bm{s}}_1^{(1)}, \tilde{\bm{s}}_2^{(1)}, \tilde{\bm{u}}_1^{(1)}, \tilde{\bm{u}}_2^{(1)}, \tilde{\bm{w}}_1^{(1)}, \tilde{\bm{w}}_2^{(1)}, \bm{X}_1^{(b)}, \bm{X}_2^{(b)}, \bm{Y}_1^{(b)}, \bm{Y}_2^{(b)})\notin T_{\epsilon}^{(n)}\}}.\\
    \scalemath{0.92}{\mathcal{E}_1^{(B+1)}}&\triangleq &\scalemath{0.92}{\{(\bm{s}_1^{(B+1)}, \bm{s}_2^{(B+1)}, \bm{u}_1^{(B+1)}, \bm{u}_2^{(B+1)}, \tilde{\bm{S}}_1^{(B+1)}, \tilde{\bm{S}}_2^{(B+1)}, \tilde{\bm{U}}_1^{(B+1)}(\hat{M}_1^{(B)}), \tilde{\bm{U}}_2^{(B+1)}(M_2^{(B)}), \tilde{\bm{W}}_1^{(B+1)}, \tilde{\bm{W}}_2^{(B+1)}}, \nonumber\\
    & & \qquad\qquad\qquad\qquad\qquad\qquad\qquad\qquad\qquad\qquad\qquad\qquad\qquad\scalemath{0.92}{\bm{X}_1^{(B+1)}, \bm{X}_2^{(B+1)}, \bm{Y}_1^{(B+1)}, \bm{Y}_2^{(B+1)})\notin T_{\epsilon}^{(n)}\}}.\IEEEeqnarraynumspace\\
    \scalemath{0.92}{\mathcal{E}_1^{(b)}}&\triangleq &\scalemath{0.92}{\{(\bm{S}_1^{(b)}, \bm{S}_2^{(b)}, \bm{U}_1^{(b)}(M_1^{(b)}), \bm{U}_2^{(b)}(M_2^{(b)}), \tilde{\bm{S}}_1^{(b)}, \tilde{\bm{S}}_2^{(b)}, \tilde{\bm{U}}_1^{(b)}(\hat{M}_1^{(b-1)}), \tilde{\bm{U}}_2^{(b)}(M_2^{(b-1)})}, \nonumber\\
& & \qquad\qquad\qquad\qquad\qquad\qquad\qquad\qquad\quad\scalemath{0.92}{\tilde{\bm{W}}_1^{(b)}, \tilde{\bm{W}}_2^{(b)}, \bm{X}_1^{(b)}, \bm{X}_2^{(b)}, \bm{Y}_1^{(b)}, \bm{Y}_2^{(b)})\notin T_{\epsilon}^{(n)}\},\text{\ for\ } b=2, 3, \dots, B.} 
  \end{IEEEeqnarray}
\end{figure*}

\noindent\underline{\emph{Codebook Generation}}: Given a configuration in $\Pi_Z(D_1, D_2)$, generate two length-$n$ sequences $(\tilde{\bm{s}}^{(1)}_1, \allowbreak\tilde{\bm{s}}^{(1)}_2, \allowbreak\tilde{\bm{u}}^{(1)}_1, \allowbreak\tilde{\bm{u}}^{(1)}_2, \allowbreak\tilde{\bm{w}}^{(1)}_1, \allowbreak\tilde{\bm{w}}^{(1)}_2)$ and $(\bm{s}^{(B+1)}_1, \allowbreak\bm{s}^{(B+1)}_2, \allowbreak\bm{u}^{(B+1)}_1, \allowbreak\bm{u}^{(B+1)}_2)$ to initialize and terminate the $(B+1)$-blocks encoding process with distributions 
\begin{IEEEeqnarray}{l}
\scalemath{0.95}{P_{\tilde{\bm{S}}^{(1)}_1, \tilde{\bm{S}}^{(1)}_2, \tilde{\bm{U}}^{(1)}_1, \tilde{\bm{U}}^{(1)}_2, \tilde{\bm{W}}^{(1)}_1, \tilde{\bm{W}}^{(1)}_2}(\tilde{\bm{s}}^{(1)}_1, \tilde{\bm{s}}^{(1)}_2, \tilde{\bm{u}}^{(1)}_1, \tilde{\bm{u}}^{(1)}_2, \tilde{\bm{w}}^{(1)}_1, \tilde{\bm{w}}^{(1)}_2)}\nonumber\\
\ \ \ \scalemath{0.95}{=\prod_{i=1}^n P_{\tilde{S}_1, \tilde{S}_2, \tilde{U}_1, \tilde{U}_2, \tilde{W}_1, \tilde{W}_2}(\tilde{s}^{(1)}_{1, i}, \tilde{s}^{(1)}_{2, i}, \tilde{u}^{(1)}_{1, i}, \tilde{u}^{(1)}_{2, i}, \tilde{w}^{(1)}_{1, i}, \tilde{w}^{(1)}_{2, i})}\nonumber
\end{IEEEeqnarray}
and
\begin{IEEEeqnarray}{l}
\scalemath{0.95}{P_{\bm{S}^{(B+1)}_1, \bm{S}^{(B+1)}_2, \bm{U}^{(B+1)}_1, \bm{U}^{(B+1)}_2}(\bm{s}^{(B+1)}_1, \bm{s}^{(B+1)}_2, \bm{u}^{(B+1)}_1, \bm{u}^{(B+1)}_2)}\nonumber\\
\ \ \ \scalemath{0.95}{=\prod_{i=1}^n P_{S_1, S_2, U_1, U_2}(s^{(B+1)}_{1, i}, s^{(B+1)}_{2, i}, u^{(B+1)}_{1, i}, u^{(B+1)}_{2, i})}. \nonumber
\end{IEEEeqnarray}
Moreover, generate codebooks $\mathcal{C}^{(b)}_j\triangleq\{\bm{U}^{(b)}_j(m^{(b)}_j): m^{(b)}_j=1, 2, \dots, 2^{\mli{nR^{(b)}_j}}\}$ for $b=1, 2, \dots, B$ and $j=1, 2$, where $\bm{U}^{(b)}_j(m^{(b)}_j)$ is a length-$n$ sequence distributed according to $P_{\bm{U}_j}(\bm{u}^{(b)}_j(m^{(b)}_j))=\allowbreak\prod_{i=1}^n P_{U_j}(u^{(b)}_{j, i}(m^{(b)}_j))$ and $\bm{U}^{(b)}_j(m^{(b)}_j)$'s are independent of each other. 
The initialization and termination sequences and all codebooks are revealed to both terminals.
We note that due to the construction of the Markov chain $\{Z^{(t)}\}$, the codebook $\mathcal{C}^{(b)}_j$ is also used for $\tilde{\bm{U}}_j^{(b+1)}$.\smallskip

\noindent\underline{\emph{Encoding}}: For $b=1, 2, \dots, B$ and $j=1, 2$, terminal~$j$ finds $m^{(b)}_j$ such that $(\bm{s}^{(b)}_j, \bm{u}(m^{(b)}_j))\in\mathcal{T}_{\epsilon_1}^{(n)}$. If there is more than one such index, the encoder chooses one of them at random. If there is no such index, it chooses an index at random from $\{1, 2, \dots, 2^{\mli{nR_j^{(b)}}}\}$. The transmitter then sends $\bm{x}^{(b)}_j$, where $x^{(b)}_{j, i}=F_j(s_{j, i}^{(b)}, u_{j, i}^{(b)}(m_j^{(b)}), \tilde{s}_{j,i}^{(b)}, \tilde{u}_{j, i}^{(b)}, \tilde{w}_{j, i}^{(b)})$ for $i=1, 2, \dots, n$, $\tilde{s}_{j,i}^{(b)}=s_{j,i}^{(b-1)}$, $\tilde{u}_{j, i}^{(b)}=u_{j, i}^{(b-1)}$, and $\tilde{w}_{j, i}^{(b)}=(x_{j, i}^{(b-1)}, y_{j, i}^{(b-1)})$.
For $b=B+1$, $\bm{x}^{(B+1)}$ is generated in the same way using the termination sequence.\smallskip

\noindent\underline{\emph{Decoding}}: For $b{=}2, 3, \dots, B+1$ and $j, j'{=}1, 2$ with $j{\neq} j'$, terminal~$j$ finds an index $\hat{m}^{(b-1)}_{j'}$ such that 
$(\bm{s}_j^{(b)}, \allowbreak\bm{u}_j^{(b)}, \allowbreak\tilde{\bm{s}}_j^{(b)}, \allowbreak\tilde{\bm{u}}_j^{(b)}, \allowbreak\tilde{\bm{u}}_{j'}^{(b)}(\hat{m}^{(b-1)}_{j'}), \allowbreak\tilde{\bm{w}}^{(b)}_j, \bm{x}^{(b)}_j, \allowbreak\bm{y}^{(b)}_j)\in\mathcal{T}^{(n)}_{\epsilon},$    
where $\tilde{\bm{u}}_{j'}^{(b)}(\hat{m}^{(b-1)}_{j'})\in\mathcal{C}_{j'}^{(b-1)}$. If there is more than one choice, the decoder chooses one of them at random. 
If there is no such index, it chooses one at random from $\{1, 2, \allowbreak\dots, \allowbreak 2^{\mli{nR_{j'}^{(b)}}}\}$.
The source messages $\bm{s}_{j'}^{(b-1)}$ is then reconstructed via $\tilde{s}^{(b-1)}_{j', i}=G_j(\tilde{u}_{j', i}^{(b)}(\hat{m}^{(b-1)}_{j'}), \allowbreak s^{(b)}_{j, i}, \allowbreak u^{(b)}_{j, i}, \allowbreak \tilde{s}_{j, i}^{(b)}, \allowbreak\tilde{u}_{j, i}^{(b)}, \allowbreak\tilde{w}^{(b)}_{j, i}, \allowbreak y^{(b)}_{j, i})$ for $i=1, 2, \dots, n$.\smallskip

\noindent\underline{\emph{Performance Analysis}}:
Let $M^{(b)}_j$ and $\hat{M}^{(b)}_j$ denote the random encoded and decoded indices for $\bm{S}^{(b)}_j$. 
We first define the events $\mathcal{E}_1^{(b)}$, $b=1, 2, \dots, B+1$, in \eqref{eq:errorevent} for terminal~1. 
We analogously define the events $\mathcal{E}_2^{(b)}$ for terminal~$2$ (not shown here) and consider the error event $\mathcal{E}=\cup_{b=1}^{B+1} \mathcal{E}_1^{(b)}\cup \mathcal{E}_2^{(b)}$. 
The expected distortion of terminal~$j$'s source reconstruction (averaged with respect to all codebooks, source messages, channel inputs, and channel outputs) can be bounded by
\begin{IEEEeqnarray}{l}
\frac{1}{B}\sum_{b=1}^B\mathbb{E}[d_j(\bm{S}^{(b)}_j, \hat{\bm{S}}^{(b)}_j)]\nonumber\\
\ \ \ \ \le  \Pr(\mathcal{E}) d_{j, \max}+\frac{1}{B}\sum_{b=1}^B \Pr\big(\overline{\mathcal{E}}\big)\mathbb{E}[d_j(\bm{S}^{(b)}_j, \hat{\bm{S}}^{(b)}_j)|\overline{\mathcal{E}}]\label{eq:distanalysisCE}\IEEEeqnarraynumspace\\
\ \ \ \ \le  \Pr(\mathcal{E}) d_{j, \max}+\frac{1}{B}\sum_{b=1}^B (1+\epsilon)\mathbb{E}[d_j(S^{(b)}_j, \hat{S}^{(b)}_j)]\label{eq:distanalysisavertypical}\\
\ \ \ \ =  \Pr(\mathcal{E}) d_{j, \max}+(1+\epsilon)\mathbb{E}[d_j(S_j, \hat{S}_j)]\label{eq:distanalysisstat}\\
\ \ \ \ \le  \Pr(\mathcal{E}) d_{j, \max}+(1+\epsilon) D_j,
\end{IEEEeqnarray}
where \eqref{eq:distanalysisCE} follows from $\mathbb{E}[d_j(\bm{S}^{(b)}_j, \hat{\bm{S}}^{(b)}_j)|\mathcal{E}]\le d_{j, \max}$ with $d_{j, \max}\triangleq \max_{s_j, \hat{s}_j}d_j(s_j, \hat{s}_j)$, \eqref{eq:distanalysisavertypical} is due to the typical average lemma \cite{kim2011}, \eqref{eq:distanalysisstat} follows from the stationarity of the Markov chain, and the last inequality holds by assumption. 

If we can further show that $\Pr\big(\mathcal{E}\big)\to 0$ and the joint source-channel coding rate goes to one as both $n$ and $B$ go to infinity, then the distortion pair $((1+\epsilon)D_1, (1+\epsilon)D_2)$ is achievable. 
Note that it suffices to show that $\Pr\big(\mathcal{E}^{(1)}_j\big)\to 0$ and $\Pr\big(\mathcal{E}^{(b)}_j\cap\overline{\mathcal{E}}_j^{(b-1)}\big)\to 0$ for all $j=1, 2$ and $b=2, 3, \dots, B+1$ since 
\[
  \scalemath{0.95}{\mathcal{E}_j^{(b)}\subseteq\mathcal{E}_j^{(b-1)}\cup(\mathcal{E}_j^{(b)}\cap\overline{\mathcal{E}}_j^{(b-1)})\subseteq\mathcal{E}_j^{(1)}\cup\Bigg(\bigcup\limits_{t=2}^{b}\mathcal{E}_j^{(t)}\cap\overline{\mathcal{E}}_j^{(t-1)}\Bigg)},
\] 
where the second inclusion relationship is obtained by successive application of the first one ($b-1$ times), and hence 
\begin{IEEEeqnarray}{l}
  \scalemath{0.95}{\Pr(\mathcal{E})\le (B+1)\Bigg[\Pr(\mathcal{E}_1^{(1)})+\Pr(\mathcal{E}_2^{(1)})}\nonumber\\
  \qquad\quad\ \scalemath{0.95}{+\sum_{b=2}^{B+1}\left(\Pr(\mathcal{E}_1^{(b)}\cap\overline{\mathcal{E}}_1^{(b-1)})+\Pr(\mathcal{E}_2^{(b)}\cap\overline{\mathcal{E}}_2^{(b-1)})\right)\Bigg]}.\nonumber
\end{IEEEeqnarray}

Due to symmetry, we only analyze $\Pr\big(\mathcal{E}^{(1)}_1\big)$ and $\Pr\big(\mathcal{E}^{(b)}_1\cap\overline{\mathcal{E}}_1^{(b-1)}\big)$ for the reconstructions of $\bm{S}_1^{(b)}$ below.
For $j=1, 2$ and $b=1, 2, \dots, B+1$, we first define
\begin{IEEEeqnarray}{l}
\scalemath{0.93}{\mathcal{F}^{(b)}_j =\{(\bm{S}_j^{(b)}, \bm{U}_j^{(b)}(m_j^{(b)}))\notin\mathcal{T}^{(n)}_{\epsilon_1}\ \text{for all}\ m_j^{(b)}\}},\nonumber\vspace{+0.2cm}\\
\scalemath{0.93}{\mathcal{F}^{(b)}_3=\{(\bm{S}_1^{(b)}, \bm{S}_2^{(b)}, \bm{U}_1^{(b)}(M_1^{(b)}), \bm{U}_2^{(b)}(M_2^{(b)}), \tilde{\bm{S}}_1^{(b)}, \tilde{\bm{S}}_2^{(b)}}, \nonumber\\
\qquad\qquad\qquad \ \ \scalemath{0.93}{\tilde{\bm{U}}_1^{(b)}(M_1^{(b-1)}), \tilde{\bm{U}}_2^{(b)}(M_2^{(b-1)}), \tilde{\bm{W}}_1^{(b)}, \tilde{\bm{W}}_2^{(b)}}, \nonumber\\
\qquad\qquad\qquad\qquad\qquad\ \ \ \ \ \ \scalemath{0.93}{\bm{X}_1^{(b)}, \bm{X}_2^{(b)}, \bm{Y}_1^{(b)}, \bm{Y}_2^{(b)})\notin\mathcal{T}^{(n)}_{\epsilon}\}},\nonumber\vspace{+1cm}\\ 
\scalemath{0.93}{\mathcal{F}^{(b)}_4 =\{\exists\ \hat{m}^{(b-1)}_1\neq M^{(b-1)}_1\ \text{s.t.}\ (\bm{S}_2^{(b)}, \bm{U}_2^{(b)}(M_2^{(b)}), \tilde{\bm{S}}_2^{(b)}}, \nonumber\\  
\qquad\ \ \scalemath{0.93}{\tilde{\bm{U}}_1^{(b)}(\hat{m}_1^{(b-1)}), \tilde{\bm{U}}_2^{(b)}(M_2^{(b-1)}), \tilde{\bm{W}}_2^{(b)}, \bm{X}_2^{(b)}, \bm{Y}_2^{(b)})\in\mathcal{T}^{(n)}_{\epsilon}\}},\nonumber
\end{IEEEeqnarray}
with the exceptions that $\mathcal{F}^{(1)}_3\triangleq\mathcal{E}_1^{(1)}$ and $\mathcal{F}^{(B+1)}_3\triangleq\mathcal{E}_1^{(B+1)}$ due to the initialization and termination phases of the encoding process. 
Next, we use the following results to obtain \eqref{eq:mainconda}. \medskip

\noindent {\it Claim~1}: For $b=2, 3, \dots, B+1$, the event $\overline{\mathcal{F}}^{(b)}_3\cap\overline{\mathcal{F}}^{(b)}_4$ implies that $\hat{M}_1^{(b-1)}=M_1^{(b-1)}$.\smallskip 

\noindent {\it Claim~2}: $\mathcal{E}_1^{(1)}\subseteq \mathcal{F}^{(1)}_1\cup \mathcal{F}^{(1)}_2\cup (\overline{\mathcal{F}}^{(1)}_1\cap\overline{\mathcal{F}}^{(1)}_2\cap\mathcal{E}_1^{(1)})$\smallskip

\noindent {\it Claim~3}: $\mathcal{E}_1^{(B+1)}\cap\overline{\mathcal{E}}_1^{(B)}\subseteq (\mathcal{F}_3^{(B+1)}\cap\overline{\mathcal{E}}_1^{(B)}) \cup \mathcal{F}_4^{(B+1)}$\smallskip

\noindent {\it Claim~4}: The relationship: $\mathcal{E}_1^{(b)}\cap\overline{\mathcal{E}}_1^{(b-1)}\subseteq \mathcal{F}^{(b)}_1\cup \mathcal{F}^{(b)}_2\cup (\overline{\mathcal{F}}^{(1)}_1\cap\overline{\mathcal{F}}^{(1)}_2\cap\mathcal{F}_3^{(b)}\cap\overline{\mathcal{E}}_1^{(b-1)})\cup \mathcal{F}_4^{(b)}$ holds for $b=2, 3, \dots, B$. \smallskip

\noindent {\it Claim~5}: If $R^{(1)}_j>I(S_j; U_j)+\delta_1(\epsilon_1)$ for $j=1, 2$, then $\lim_{n\to\infty}\Pr\big(\mathcal{E}_j^{(1)}\big)=0$.\smallskip

\noindent {\it Claim~6}: If $R^{(B)}_1<I(\tilde{U}_1; S_{2}, U_{2}, \tilde{S}_{2}, \tilde{U}_{2}, \tilde{W}_{2}, X_{2}, Y_{2})-\delta(\epsilon)$, then $\lim_{n\to\infty}\Pr\big(\mathcal{E}_1^{(B+1)}\cap\overline{\mathcal{E}}_1^{(B)}\big)=0$.\smallskip

\noindent {\it Claim~7}: For $b=2, 3, \dots, B$, if $R^{(b)}_j > I(S_j; U_j)+\delta_1(\epsilon_1)$ for $j=1, 2$ and $R^{(b-1)}_1<I(\tilde{U}_1; S_{2}, U_{2}, \tilde{S}_{2}, \allowbreak\tilde{U}_{2}, \tilde{W}_{2}, X_{2}, Y_{2})-\delta(\epsilon)$, then $\lim_{n\to\infty}\Pr\big(\mathcal{E}_{1}^{(b)}\cap\overline{\mathcal{E}}_{1}^{(b-1)}\big)=0$.\medskip

The non-negative quantities $\delta_1(\epsilon_1)$ and $\delta(\epsilon)$ above arise from the standard typicality arguments and $\lim_{\epsilon_1\to 0}\delta_1(\epsilon_1)=0$ and $\lim_{\epsilon\to 0}\delta(\epsilon)=0$. 
Claims~3 and~4 are derived using the fact that $\mathcal{E}_1^{(b)}\subseteq\mathcal{F}^{(b)}_3\cup\mathcal{F}^{(b)}_4$, which is a consequence of Claim~1. 
Claims~5-7 are derived based on Claims~2-4, respectively. 
More specifically, the union bound is applied to each inclusion relationship (in Claims~2-4) to upper bound the probability of the event on the left-hand-side. 
A thorough analysis next yields the conditions in Claims~5-7, which ensure that all terms in the upper bound asymptotically vanish.
The proofs of Claims~5-7 invoke the covering lemma \cite{kim2011}, the conditional typical lemma \cite{kim2011}, and \cite[Lemma 1]{kim2015}.  

Swapping the role of terminals~1 and~2, we obtain the analogous results $\lim_{n\to\infty}\Pr\big(\mathcal{E}_2^{(1)}\big)=0$ and $\lim_{n\to\infty}\Pr\big(\mathcal{E}_{2}^{(b)}\cap\overline{\mathcal{E}}_{2}^{(b-1)}\big)=0$ for $b=2, 3, \dots, B+1$ provided that $R^{(b)}_j>I(S_j; U_j)+\delta_1(\epsilon_1)$
for $j=1, 2$ and $b=1, 2, \dots, B$ and $R^{(b-1)}_2<\allowbreak I(\tilde{U}_2; S_{1}, U_{1}, \allowbreak\tilde{S}_{1}, \allowbreak\tilde{U}_{1}, \allowbreak\tilde{W}_{1}, \allowbreak X_{1}, \allowbreak Y_{1})-\delta(\epsilon)$
for $b=2, 3, \dots, B+1$. 
Combining all conditions above then gives the two inequalities in \eqref{eq:mainconds}. 
To complete the proof, we first increase $B$ so that the JSCC rate $B/(B+1)$ is close to one. 
Fixing this choice of $B$, we next make $n$ sufficiently large to ensure that all joint typicality requirements behind Claims~5-7 (and similar claims for terminal~2) are satisfied. 
As now we have $\lim_{n\to\infty}\Pr(\mathcal{E}){=}0$ (provided that all conditions hold) and $\epsilon$ is arbitrary, the distortion pair $(D_1, D_2)$ is achievable. \hfill \IEEEQEDhere
\end{appendix}

\newpage
\balance
\bibliographystyle{IEEEtran}
\bibliography{literatureDB}

\end{document}